\documentclass[twocolumn]{svjour3}          
\smartqed  
\usepackage{aasmacro}
\usepackage{mathptmx}      
\usepackage{amsmath}

\usepackage{graphicx}
\usepackage{graphbox}
\usepackage{tikz}
\usepackage{float}
\usepackage{multirow}
\usepackage{color}

\providecommand{\keywords}[1]
{
  \small	
  \textbf{\textit{Keywords---}} #1
}

\begin{document}
	\title{Disentangling Atmospheric Compositions of K2-18 b with Next Generation Facilities}
	\subtitle{}
	\author{Quentin Changeat$^1$, Billy Edwards$^1$, Ahmed F. Al-Refaie$^1$, Angelos Tsiaras$^1$, Ingo P. Waldmann$^1$, Giovanna Tinetti$^1$}
	\institute{Corresponding author: Quentin Changeat\\ \email{quentin.changeat.18@ucl.ac.uk}  \at
              $^1$Department of Physics and Astronomy, University College London, London, UK.          
}

\date{Accepted in Experimental Astronomy: 20 August 2021}

\maketitle

\begin{abstract}

Recent analysis of the planet K2-18\,b has shown the presence of water vapour in its atmosphere. While the H$_2$O detection is significant, the Hubble Space Telescope (HST) WFC3 spectrum suggests three possible solutions of very different nature which can equally match the data. The three solutions are a primary cloudy atmosphere with traces of water vapour (cloudy sub-Neptune), a secondary atmosphere with a substantial amount (up to 50\% Volume Mixing Ratio) of H$_2$O (icy/water world) and/or an undetectable gas such as N$_2$ (super-Earth). Additionally, the atmospheric pressure and the possible presence of a liquid/solid surface cannot be investigated with currently available observations.

In this paper we used the best fit parameters from \cite{Tsiaras_k2-18} to build James Webb Space Telescope (JWST) and Ariel simulations of the three scenarios. We have investigated 18 retrieval cases, which encompass the three scenarios and different observational strategies with the two observatories. Retrieval results show that twenty combined transits should be enough for the Ariel mission to disentangle the three scenarios, while JWST would require only two transits if combining NIRISS and NIRSpec data. This makes K2-18\,b an ideal target for atmospheric follow-ups by both facilities and highlights the capabilities of the next generation of space-based infrared observatories to provide a complete picture of low mass planets. \vspace{5mm}

\end{abstract}


\section{INTRODUCTION}

Despite biases in observational techniques towards large, gaseous giant planets, current  statistics from over 4000 confirmed planets show a very different picture: low radius planets are the most abundant exoplanets, especially around late-type stars \cite{Howard2016,Dressing2013,2017_fulton,fulton2018,Dressing2017}. The frequency of these planets seems to follow a bimodal distribution when plotted against the planetary size \cite{fulton2018}, with most planets clustering around two peaks at $R_p \sim 1.3 \, R_{\oplus}$ and  $R_p \sim 2.4 \, R_{\oplus}$. A dependence of said distribution on stellar type was recently reported, with the radius gap shifting from 1.7-2 $R_{\oplus}$ for sun like stars to 1.4-1.7 $R_{\oplus}$ for K and M types \cite{fulton2018,Cloutier_2020-valley,cloutier2020toi1235}.  This observational evidence can be explained by a combination of  formation and evolution processes, but the details of these processes are still not completely understood. In planets larger than $1.7\,R_{\oplus}$ volatiles are expected to contribute significantly to the planetary composition, although it is often difficult to extract the exact bulk composition out of the knowledge of planetary radius and/or mass   (e.g. \cite{Valencia_2013}).  From a formation perspective,  in-situ formation of small-size planets is theoretically possible, but it may happen only under very specific conditions (e.g. \cite{Ikoma2012,Ogihara2015}). 
Low mass planets could also be the remnants of larger planets which have lost part of their initial gaseous envelope, due to XUV-driven hydrogen mass-loss coupled with planetary thermal evolution (e.g. \cite{Leitzinger2011,Owen2012,Lopez2012,Owen2013,Owen2017,Owen_2019}). Direct observations of the atmospheric composition may help to remove some of the degeneracies associated with  the bulk composition and nature of these planets   (e.g. \cite{Valencia_2013,Zeng_2018}), and therefore provide additional constraints to the formation and evolution scenarios currently considered in the literature 
\cite{Kite_2020,Owen2013,Owen2017}.

While the rapid development of exoplanet studies has revealed more and more about their atmospheric properties \cite{tsiaras_55cnce,Sing_2015,Line_2016,Oreshenko_2017,Barstow_10HJ,Tsiaras_pop_study_trans,evans_w121_2008,Pinhas_ten_HJ_clouds,edwards2020ares,Anisman_2020,Changeat_2020_k11,Yip_2021,Changeat_2021_k9,Changeat_2021_w43}, the study of small planets in, and around, the radius gap has so far remained very limited. 
Current observations of the atmospheres of these small worlds have not yet allowed us to infer precise constraints on their nature and only a few planets have had their atmospheres investigated. The sub-Neptune  GJ-1214\,b was observed multiple times with the Hubble Space Telescope (HST) \cite{Kreidberg_GJ1214b_clouds} and other instruments. Current observations are suggestive of a cloudy world, which have motivated an extensive literature \cite{Menou_2011,Haghighipour_2013,Valencia_2013,Kataria_2014,Charnay_2015,Gao_2018,Lavvas_2019,Miguel_2019,2019_Venot,Bourgalais_2020,Guilluy2021}. Based on its mass and radius, we speculate its atmosphere being hydrogen dominated, however so far this could not be confirmed by the direct measurement of its atmosphere. Observations of 55-Cancri\,e, an extremely hot and irradiated low mass planet with a radius $\le$ 2 $R_{\oplus}$, suggest the presence of a volatile-rich atmosphere \cite{tsiaras_55cnce,Demory_2016,Hammond_2017,Angelo_2017}. However, further observations are needed to constrain current models of the atmospheric composition and stability \cite{Ito_2015,Hammond_2017,Modirrousta_Galian_2020,Ito2021}. Hubble observations of the TRAPPIST-1 planetary system \cite{de_Wit_2016_trappist,de_Wit_2018_trappist} did not reveal the atmospheric composition  of these rocky worlds 
\cite{de_Wit_2016_trappist,de_Wit_2018_trappist,Ducrot_2018}. Recently, another sub-Neptune planet, $\pi$\,Men\,c has been proposed to host a volatile-rich atmosphere \cite{Garcia_2021}. Similarly, the temperate super-Earth LHS-1140\,b \cite{Dittman_2017} was observed by HST. These observations hinted at the presence of water vapour \cite{edwards_2021_LHS} but with a low significance. GJ-1132\,b, another super-Earth was found to host a light secondary atmospheres \cite{Swain_2021_GJ1132}, traced by the detection of features from aerosol scattering, HCN and CH$_4$. However, two independent studies \cite{Mugnai_2021_GJ1132,libby_roberts_gj1132} analysing the same dataset found a featureless spectrum for this world. For all those planets, further observations are required to remove the remaining degeneracies and truly understand the nature of these worlds. Future observatories, such as JWST, Twinkle and Ariel, are needed to provide adequate observational constraints to the modelling effort inspired by these planets \cite{Kislyakova_2017,Papaloizou_2017,Barr_2018,Grimm_2018,suissa_2018trappist1e,unterborn_2018updated,Dorn_2018,Wright_2018,Moran_2018,bolmont2018constraining,Hay_2019,Dobos_2019,Dencs_2019,Schoonenberg_2019,yang2019transition,Vecchio_2020,Hu_2020,Hori_2020}.

Recent observations of K2-18\,b with HST have, for the first time, revealed the presence of water vapour in the atmosphere of a low  mass planet (2.6R$_\oplus$, 8.6M$_\oplus$) orbiting within the habitable-zone of its star. This detection, published by two independent studies, \cite{Tsiaras_k2-18,Benneke_k2-18}, is particularly exciting if compared with the featureless atmospheric signals observed so far in the Super-Earth/Sub-Neptune regime \cite{Kreidberg_GJ1214b_clouds,de_Wit_2018_trappist}. 
Nevertheless, while the water vapour feature is evident in the HST-WFC3 observations, it is not possible to constrain its abundance, in particular relative to H/He and other undetectable gases. The narrow wavelength coverage of the HST-WFC3 camera does not allow us to distinguish between a primary atmosphere, i.e. mainly composed of H/He, or a more secondary atmosphere, i.e. an atmosphere which has evolved from the primordial composition and contains a non negligible fraction of gases other than H/He. In the future, more planets in the low mass regime like K2-18\,b will be observed using Ariel \cite{Tinetti_2021_redbook}, JWST \cite{Greene_2016_JWST} and Twinkle \cite{Edwards_twinkle}. A prime goal of these observations will be to reveal their profound nature and the way they form. In order to understand the ability of future telescopes to answer these questions, we use K2-18\,b as a prime example of this class of planets and base our simulations scenarios on its properties. 

To capture the variety of possible cases which could explain current WFC3 observations,  three main scenarios, of very different nature, were identified in \cite{Tsiaras_k2-18}. These scenarios most likely only represent a subset of the possible atmospheric composition for K2-18\,b but for simplicity, we limit our study to these three cases. These are summarised here: 
\begin{enumerate}
\item Icy/Water world: A clear secondary atmosphere with a  mean molecular weight explained by   water vapour (up to 50\% in Volume Mixing Ratio) additionally to H/He. 
\item Super-Earth: A clear secondary atmosphere with traces of water vapour and a mean molecular weight increased by one or multiple undetectable absorbers  (e.g. N$_2$) and H/He. 

\item Cloudy sub-Neptune: A cloudy primary atmosphere composed mainly by H/He with a mean molecular weight of 2.3, and traces of water vapour. 
\end{enumerate}

Most importantly, the thickness of the atmosphere cannot be inferred from the HST-WFC3 observations. This information is critical to constrain the bulk nature of the planet, i.e. whether K2-18\,b is an Ocean planet with a liquid surface or there is a thick H/He atmosphere. Simulations by \cite{scheucher_k218} suggest that K2-18\,b has an H2-He atmosphere with limited amounts of H$_2$O and CH$_4$. Their 1D climate disequilibrium-chemistry models do not support the possibility of K2-18\,b having a water reservoir directly exposed to the atmosphere. However, work by \cite{madhu_k218} showed that the constraints on the interior allow for multiple scenarios between a rocky world with massive H/He envelope to a water world with thin envelope. Alternative interpretations of the data also indicated that CH$_4$ could contribute to the observed 1.4$\mu$m feature, though with a lower Bayesian evidence than the water case \cite{Blain_2021}. The true nature of planets in this regime is currently unknown although many models have sought to use their bulk proprieties to understand them \cite{Seager_2007,Valencia_2010,Rogers_2011,Valencia_2013,Lopez_2014}.

In this paper, we simulate the ability of the European Space Agency Ariel mission to observe K2-18\,b's as an example of super-Earth/sub-Neptune atmospheres. This planet is a challenging target for Ariel, which is not designed to specifically observe the thick atmosphere of small worlds. Here we are essentially testing the limits of the Ariel Space telescope. Since the James Webb Space Telescope (JWST) will be adapted for these type of planets and could bring significant help in constraining the nature of these sub-Neptunes, we also include simulations for this telescope. We use spectral retrieval models to interpret the simulated observations and discuss future prospects to understand the nature of these worlds in light of our simulations and break the current degeneracies on their interiors.

\section{METHODOLOGY}
\label{method}

\subsection{Overview}

To simulate various chemical compositions and structures of the atmosphere of  K2-18\,b and conclude on their detectability, we performed both forward radiative transfer models and inverse models (spectral  retrievals) using the open-source Bayesian framework  TauREx3 \cite{al_refaie_taurex3}, which is a more efficient and comprehensive version of TauRex \cite{Waldmann_taurex2,Waldmann_taurex1}. TauREx is a fully Bayesian radiative transfer code  which includes the highly accurate molecular line-lists from the ExoMol project \cite{Tennyson_exomol}, HITEMP \cite{rothman}  and  HITRAN  \cite{gordon}. The  complete list of opacities used in this paper can be found in Table \ref{tab:linelists}. TauREx3 is available on GitHub \footnote{ http://github.com/ucl-exoplanets/Taurex3\_public} and is optimised for Windows, Mac and Linux. It has been benchmarked against other retrieval codes from the community \cite{Barstow_2020_bech}.

\begin{table}
\centering

$\begin{array}{|c|c|}
\mbox{Opacity} & \mbox{References}  \\
\hline
\mbox{H}_2\mbox{-H}_2 & \mbox{\cite{abel_h2-h2}, \cite{fletcher_h2-h2}} \\
\mbox{H}_2\mbox{-He} & \mbox{\cite{abel_h2-he}} \\
\mbox{H}_2\mbox{O} & \mbox{\cite{barton_h2o}, \cite{polyansky_h2o}} \\

\end{array}$
\caption{List of opacities used in this work}
\label{tab:linelists}
\end{table}  

We followed a three-step approach. We started by simulating the three scenarios described in \cite{Tsiaras_k2-18}, i.e.  a secondary atmosphere with comparable amount of H/He and water vapour,  a secondary atmosphere with  comparable amount of H/He and N$_2$ and traces of water vapour, a primary H/He  atmosphere with clouds and traces of water vapour. Since the nature of K2-18\,b is still poorly understood, we focus here on a limited number of simplified scenarios and do not consider complex atmospheric processes (self consistent chemistry, micro-physical clouds and radiative equilibrium models). The parameters used in these forward models are detailed in Table \ref{params}.

\begin{table}
\centering
$\begin{array}{|c|c|c|c|c|c|}
\mbox{Parameter} & \mbox{Scenario 1} & \mbox{Scenario 2} & \mbox{Scenario 3} & \mbox{Mode} & \mbox{Priors} \\
\hline
\mbox{radius (R$_J$)} & 0.219 & 0.219 & 0.216 & \mbox{linear} & 0.01 - 0.5\\
\mbox{T (K)} & 286 & 286 & 288 & \mbox{linear} & 50 - 600\\
\mbox{H$_2$O/H$_2$} & 0.541 & 3.71\times 10^{-4} & 1.28\times 10^{-3} & \mbox{log} & 10^{-5} - 10 \\
\mbox{N$_2$/H$_2$} & 7.82\times 10^{-7} & 0.0592 & 6.74\times 10^{-7} & \mbox{log} & 10^{-5} - 10\\
\mbox{P$_{clouds}$ (bar)} & 2.85 & 2.85 & 6.92 \times 10^{-2} & \mbox{log} & 10^{1} - 10^{-5}\\

\end{array}$

\caption{Parameters adopted to describe the tree atmospheric scenarios described in \cite{Tsiaras_k2-18} and their priors in the retrievals. The radius is expressed in Jupiter radius (R$_J$), the temperature (T) in Kelvin and the abundances are expressed in Volume Mixing ratios. The parameter P$_{clouds}$ corresponds to the top pressure of our Grey cloud deck. From the best-fit models in \cite{Tsiaras_k2-18}, we only keep three significant figures, which can lead to small differences in our simulations.}  Scenario 1: secondary atmosphere with comparable amount of H/He and H$_2$O. Scenario 2:  secondary atmosphere with  comparable amount of H/He and N$_2$ and traces of H$_2$O. Scenario 3:  primary H/He  atmosphere with clouds and traces of H$_2$O.
\label{params}
\end{table}  

Transit spectra were generated with TauREx3 at high resolution and then binned to the resolution of the observations. To simulate JWST and Ariel performances, we used the noise simulators ExoWebb, an adapted version of the tool described in \cite{Exowebb}, for JWST and ArielRad \cite{mugnai_Arielrad} for Ariel. For most of this work, we assumed the JWST observations are performed with NIRISS and NIRSpec, therefore the total number of observations reported here should be interpreted as  equally split between these two instruments. The combination of NIRISS and NIRSpec ensure a wavelength coverage from 0.8$\mu$m to 5$\mu$m, which best matches the Ariel wavelength coverage, allowing for an adequate comparison. In  the Appendix, Figure \ref{fig:jwst_inst}, we investigate other configurations for JWST (NIRISS + NIRSpec, NIRISS only, NIRSpec only or MIRI only) on the water scenario. Our results indicate that JWST may be able to provide adequate results using NIRISS only, for this particular star and planet. The use of NIRSpec or MIRI alone may not provide the best performances. We also note that adding MIRI to the NIRISS + NIRSpec setup does not bring further constraints for the scenarios investigated here. The instrument setup chosen is summarised in Table \ref{jwst_setup}. We convolved the high resolution spectra from TauREx with the instrument profiles of JWST and Ariel at native resolution. The raw spectra (taken at the focal plane native resolution) are then binned to reach a higher SNR on the data points as the spectral features of interest in this range are broad. When performing the retrievals, we do not draw a scattered instance of our raw simulated observed spectra for the reasons outlined in \cite{Feng_retrieval_earthanalog,Changeat_2layer,Changeat_2020}.

\begin{table}[]
    \centering
    \resizebox{0.45\textwidth}{!}{%
    \begin{tabular}{llll} \hline\hline
       Parameter  & NIRISS & NIRSpec  & MIRI\\ \hline
       Filter/Grism  & GR700XD ORD1 & F290LP-G395m & P750L \\
       Spectral Coverage [$\mu$m] & 0.83 - 2.81 & 2.87 - 5.27 & 5 - 12 \\
       Number Groups & 31 & 25 & 78\\
       Exposure Time [s] & 164.7 & 21.7 & 12.25\\
       Max Saturation Level & 78\% & 78\% & 79\% \\
       In Transit Integrations & 58 & 430 & 830 \\
       Out Transit Integrations & 116 & 860 & 1660 \\ \hline \hline
    \end{tabular}
    }
    \caption{JWST instrument setups used in ExoWebb for this work which utilises the latest version of Pandeia \cite{Pontoppidan}.}
    \label{jwst_setup}
\end{table}{}

The simulated observed spectra were then used as input to TauREx3, retrieval mode,  to analyse their information content and assess, by inspection of the posteriors, whether the three atmospheric scenarios could be disentangled. 

\subsection{Forward model assumptions}

As the nature of K2-18\,b is still poorly known,  we have adopted very basic assumptions in our forward models. As in  \cite{Tsiaras_k2-18}, we have assumed isothermal and isocompositional atmospheres;  clouds are simulated using a basic grey cloud model, where  the atmosphere is completely opaque below a given pressure. We included absorptions from H$_2$O and N$_2$, Collision Induced Absorption (CIA) opacities for H$_2$-H$_2$ and H$_2$-He and Rayleigh scattering \cite{cox_allen_rayleigh}. In this work, we do not consider more complex CIA opacities (for example H$_2$-H$_2$O or H$_2$O-N$_2$), and this should be explored in future works. N$_2$, being an inactive species, contributes only to the continuum and does not show molecular features. We expressed abundances as Volume Mixing Ratios. The list of parameters for each scenario is detailed in Table \ref{params}.

\begin{table}[]
    \centering
    \resizebox{\columnwidth}{!}{
    \begin{tabular}{cccc}\hline \hline
        Telescope & Number Transits & Final Resolving Power & Time Required [hrs]\\\hline
        JWST &  2 & 30 & 16 \\
        JWST &  10 & 100 & 80 \\
        JWST &  20 & 100 & 160 \\
        Ariel & 10 & 10/12/10 & 80 \\
        Ariel & 20 & 10/12/10 & 160 \\
        Ariel & 50 & 10/50/15 & 400\\\hline \hline
    \end{tabular}
    }
\caption{Simulations of K2-18\,b with JWST and Ariel as reported in this paper. The number of transits considered and the spectral resolutions are indicated. }
\label{instruments}
\end{table}

For both missions we investigated three cases by varying the number of observed transits. The list of investigated cases is summarised in Table \ref{instruments}. Being K2-18 a relatively faint star and K2-18\,b a small and cold planet, this target is challenging for Ariel, so a larger number of stacked transits is considered compared to JWST.
As this planet has  an orbital period of approximately thirty-two days, we restrain the maximum number of observed transits to fifty.  

\subsection{Retrieval Model Assumptions}

While some studies choose to use the SNR between models as proof of detectability \cite{wunderlich_snr,Lustig-Yaeger}, we performed a retrieval analysis to fully assess the capabilities of these future instruments.
We performed, in total, eighteen retrieval cases, i.e. 3 atmospheric scenarios $\times$ 2 observatories $\times$ 3 maximum number of combined transits. We used the nested sampling algorithm Multinest \cite{Feroz_multinest} with an evidence tolerance of 0.5 and 750 live points to fit our simulated spectra. We retrieved the following free parameters: planetary radius at 10 bar, atmospheric temperature $T_s$, water to hydrogen ratio (H$_2$O/H$_2$), nitrogen to hydrogen ratio (N$_2$/H$_2$) and cloud top pressure ($P_{clouds}$). For each of the fitted parameters, we used the same uniform priors to avoid biases in the comparison. The retrieved parameters and the priors adopted for the retrievals are listed in Table \ref{params}. The planetary mass is not retrieved as better constrained from radial velocity \cite{Changeat_2020_mass}.

\section{RESULTS}
\label{results}

\begin{center}
\begin{figure*}
\includegraphics[scale=0.5]{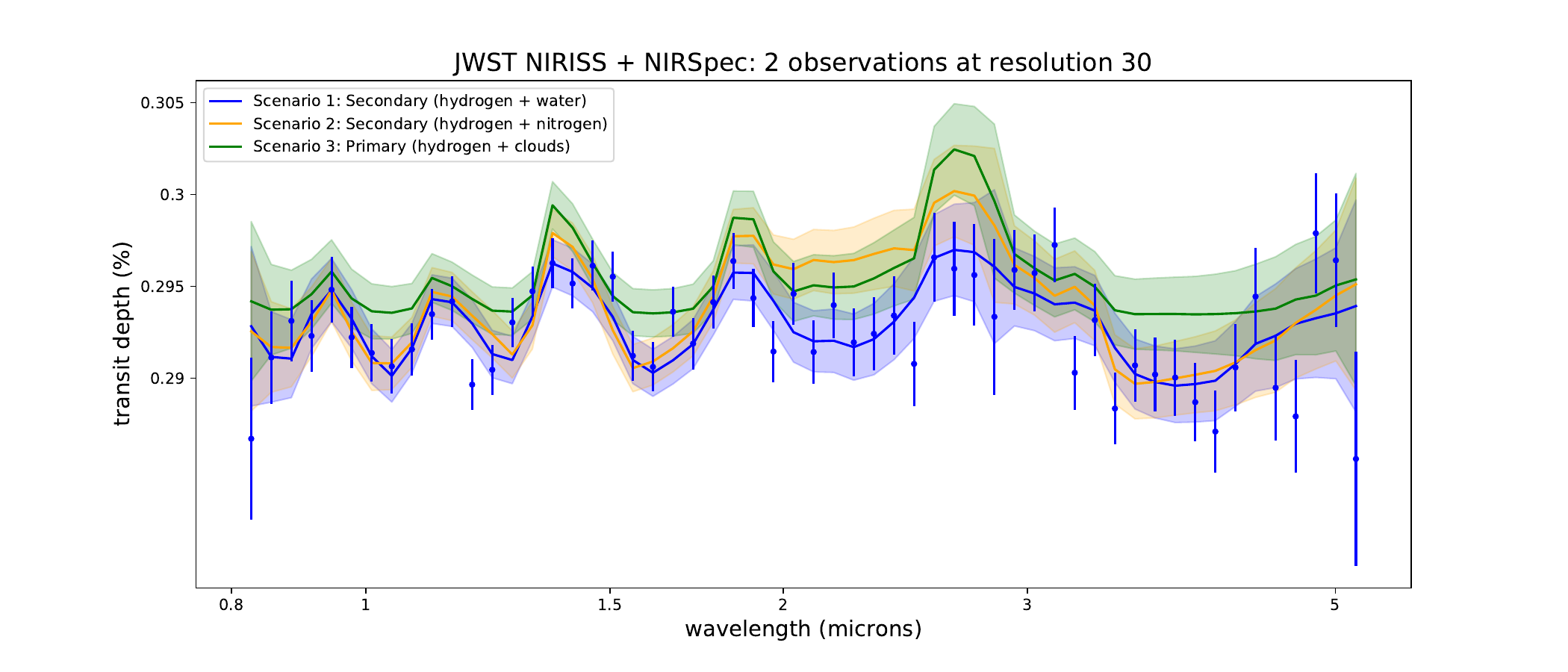}
\includegraphics[scale=0.5]{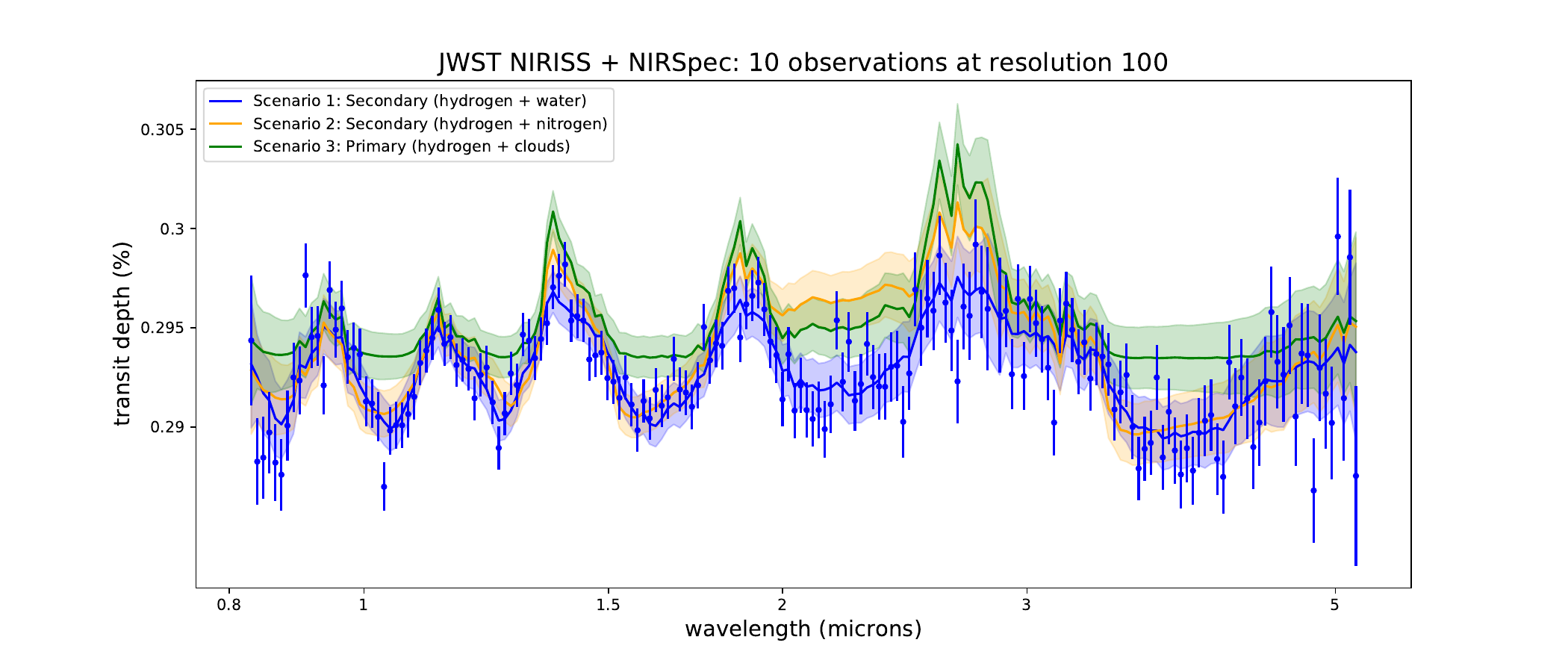}
\caption{
Simulated observed spectra for our three scenarios of K2-18b with 1$\sigma$ uncertainties obtained by combining a number of transits recorded with JWST. Top: 2 stacked transits. Bottom: 10 stacked transits. For solution 1, we also display a scattered instance of our simulated observation. Un-scattered spectra are used for the retrievals.
}\label{fig:jwst_forward}
\end{figure*}
\end{center}

\begin{center}
\begin{figure*}
\includegraphics[scale=0.5]{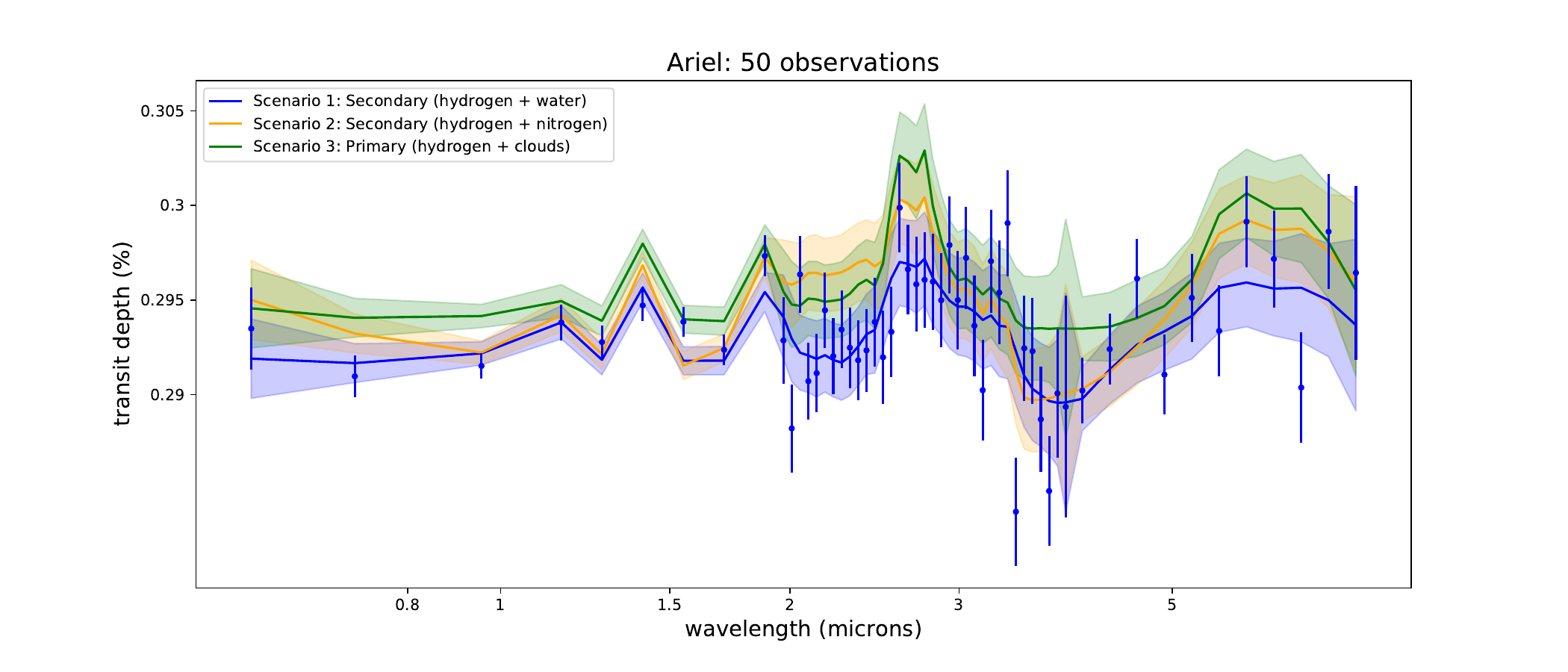}
\caption{
Simulated observed spectra for our three scenarios of K2-18b obtained by combining 50 transits recorded with Ariel. If 20 transits are combined, the three scenarios are difficult to distinguish. For solution 1, we also display a scattered instance of our simulated observation. Un-scattered spectra are used for the retrievals.
}\label{fig:ariel_forward}
\end{figure*}
\end{center}

Figure \ref{fig:jwst_forward} illustrates the case where two and ten transits  of K2-18\,b  observed with JWST NIRISS and NIRSpec are combined. Figure \ref{fig:ariel_forward} shows the simulated spectrum of K2-18\,b when fifty transits  observed with Ariel  are combined. 

\begin{center}
\begin{figure*}
\centering
\includegraphics[scale=0.4]{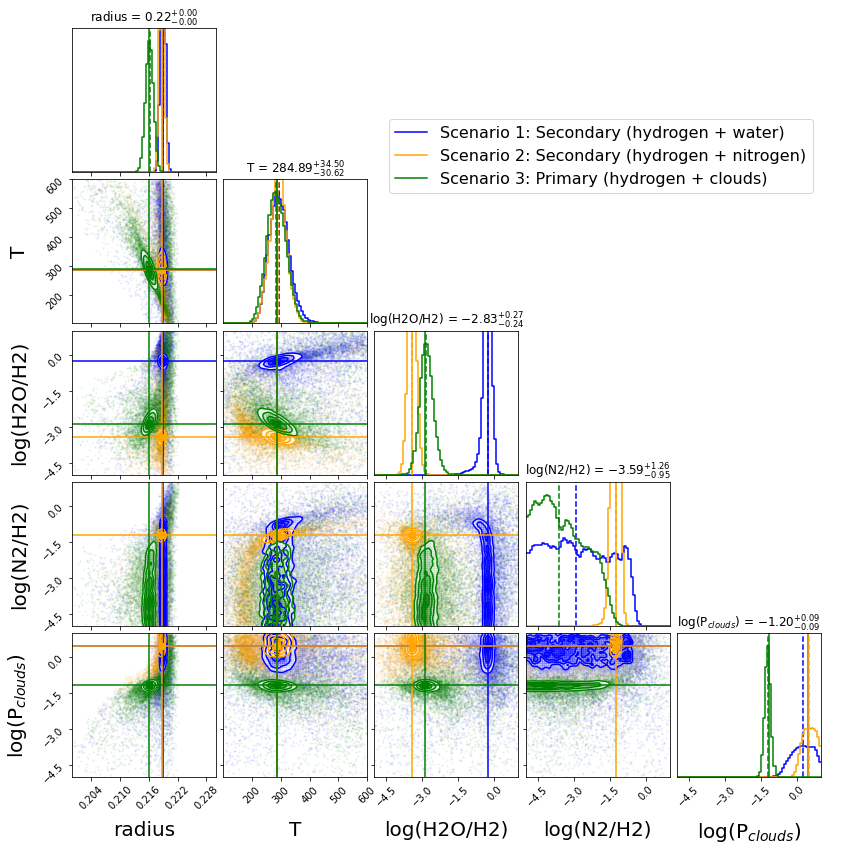}
\caption{
Posteriors related to the three atmospheric scenarios for 10 combined transits recorded with JWST. The three scenarios can be easily  distinguished by inspection of the posterior distributions of the parameters.
}\label{fig:scenario1}
\end{figure*}
\end{center}

In Figure \ref{fig:scenario1} we show the posterior distributions for two particular cases: JWST 10 transits with NIRISS and NIRSpec and Ariel 50 transits. We find that JWST and Ariel will be able to distinguish among the three scenarios presented in \cite{Tsiaras_k2-18}.   This is also true in general for the other investigated cases and we report the results of our retrievals in the Appendix. 

\begin{center}
\begin{figure*}
\centering
\includegraphics[scale=0.4]{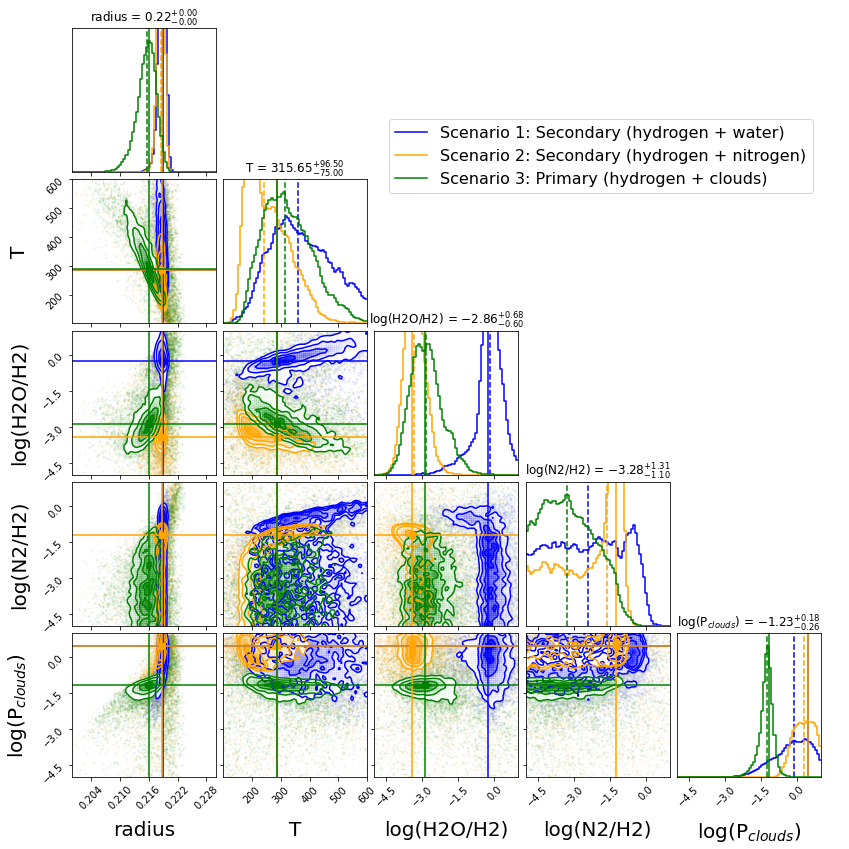}
\caption{
Posteriors related to the three atmospheric scenarios for 50 combined transits recorded with Ariel. The three scenarios can be easily  distinguished by inspection of the posterior distributions of the parameters.
}\label{fig:scenario1b}
\end{figure*}
\end{center}

K2-18\,b is a small and cold planet orbiting a faint star, therefore it is optimal for JWST sensitivity, but expected to be challenging for Ariel which is optimised for warm and hot planets around bright stars \cite{edwards_ariel}. According to our simulations, with 1 combined NIRISS and NIRSpec observation (2 transits required), JWST should be able to inform on the nature of K2-18\,b. Ariel can also reach the same conclusions, but it will require more observations: while 10 combined observations with Ariel start to  indicate  the atmospheric differences (see the Appendix), it is only after 20 combined transits that we distinguish among the three scenarios. 50 combined transits would provide a real insight on the atmospheric composition: this plan would require to observe all transits available during the nominal and extended mission lifetime. Given the fact that such plan would certainly affect the rest of the Ariel program and that JWST will likely observe K2-18\,b first, such a strategy is unlikely. This scenario, however, provides an idea of Ariel capabilities for small planets. 

In all the cases analysed here,  the water abundance is always well retrieved due to the strong molecular features. The radius at 10 bar is always very well constrained -- less than 2 percent in the worst case of Ariel observing 10 combined transits--. Cloud parameters are  also retrieved correctly in all the simulations, with very small uncertainties.
The temperature, however, is  accurately constrained only  when we combine  more than 10 JWST transits, while all other cases do not converge to the correct solutions and have large uncertainties in the retrieved temperature.

In the case of a secondary atmosphere with  water present only as a trace gas (Scenario 2), the ratio N$_2$/H$_2$ is  retrieved correctly.  By contrast, the retrievals  provide only an upper limit for the N$_2$ abundance  in the 2 other scenarios, i.e. secondary atmosphere with mainly H$_2$O and primary cloudy atmosphere.

20 combined JWST transits offer very accurate and precise posterior distributions, allowing for an unambiguous characterisation of the atmospheric main   gases in K2-18\,b, the atmospheric temperature and cloud parameters.

\begin{center}
\begin{figure}[h]
\includegraphics[scale=0.37]{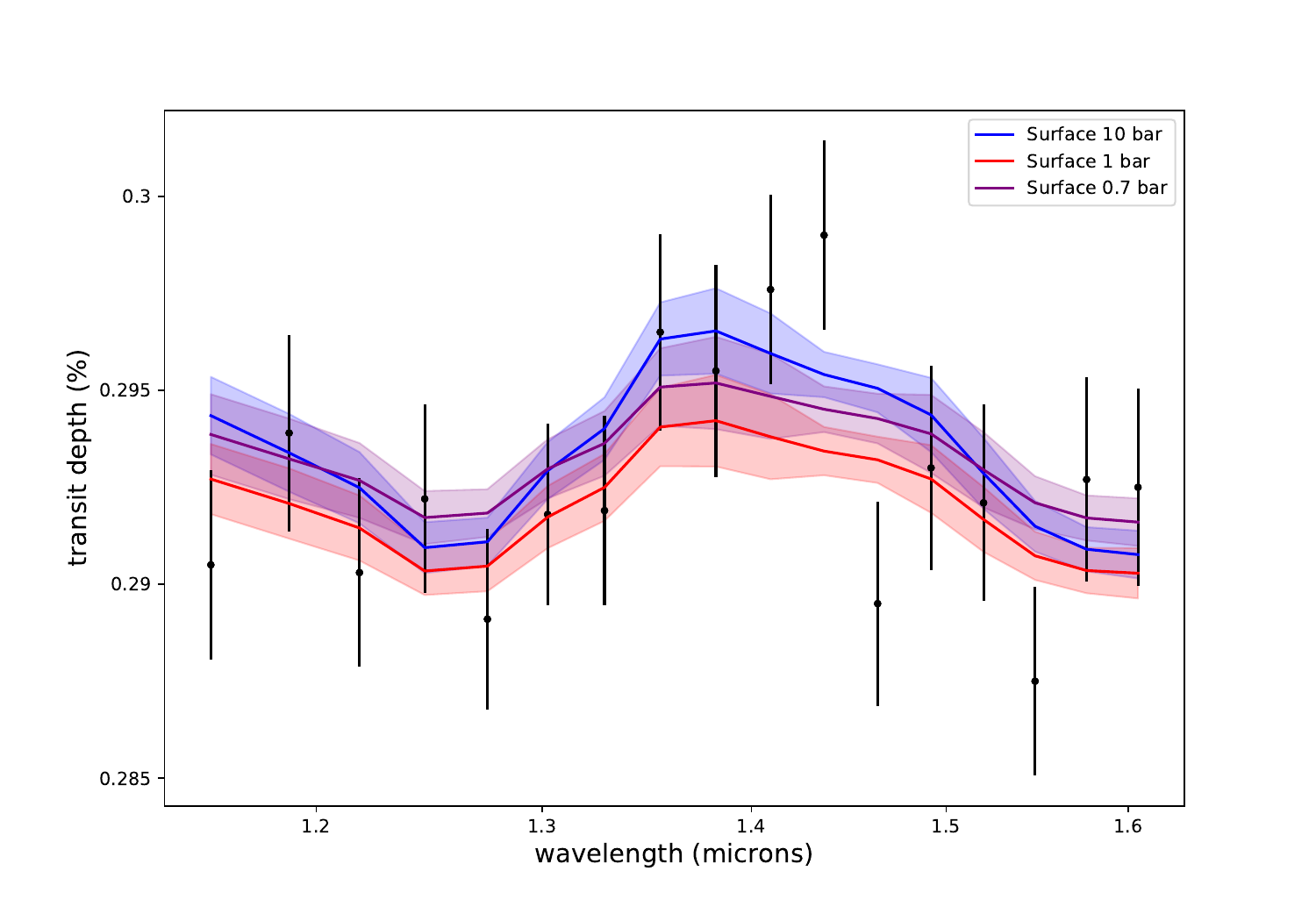}
\caption{
Best fit spectra from retrievals assuming different surface pressures to interpret the HST-WFC3 observations (black) as published in \cite{Tsiaras_k2-18}. Blue plot: 10 bar; red plot: 1 bar; purple plot: 0.7 bar.
}\label{fig:spect_pressure}
\end{figure}
\end{center}

\begin{center}
\begin{figure}[h]
\includegraphics[scale=0.37]{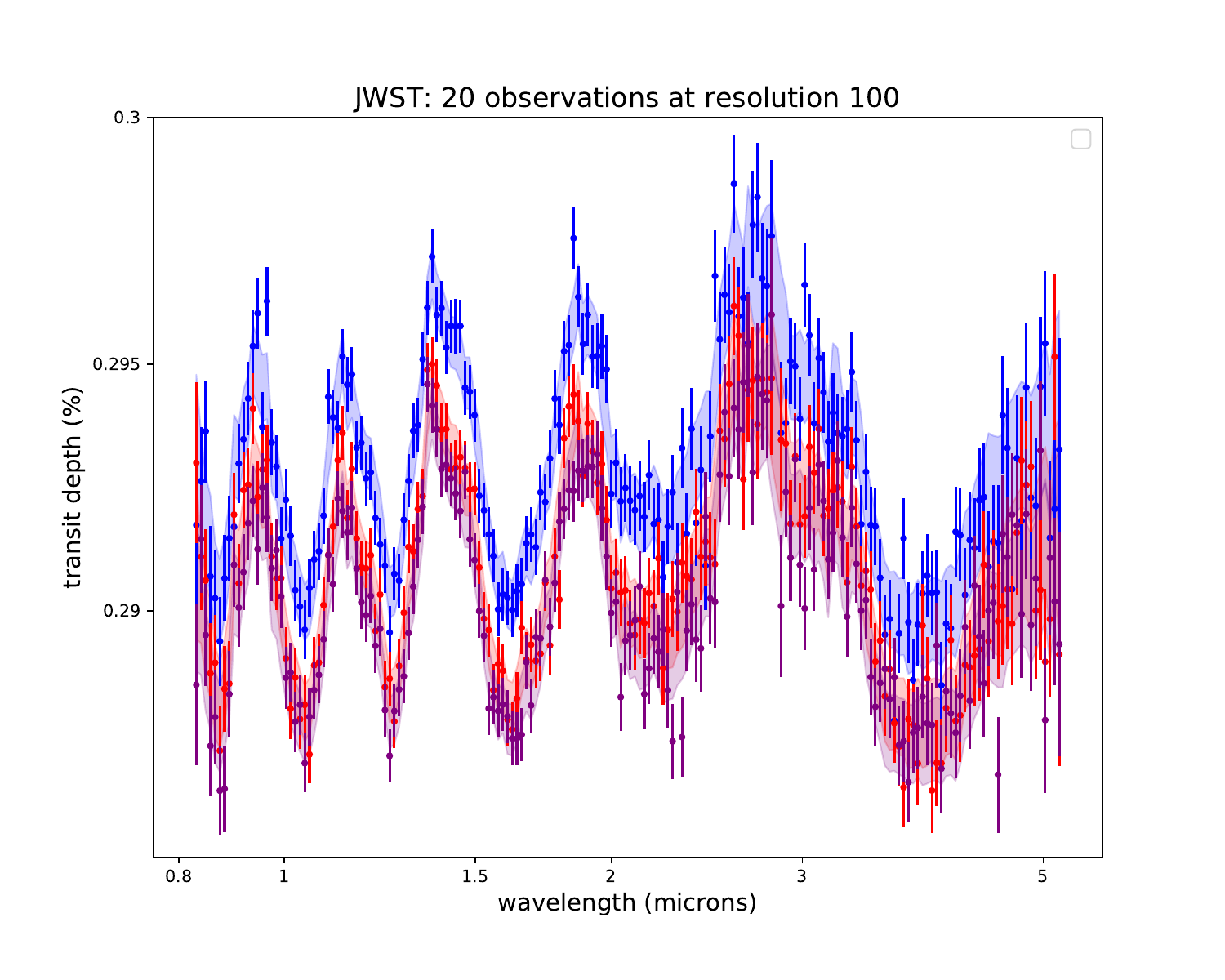}
\caption{
Simulated forward spectra assuming different surface pressures for JWST simulated observations (20 combined transits). Blue plot: 10 bar; Red plot: 1 bar; Purple plot: 0.7 bar. Un-scattered spectra are used for the retrievals.
}\label{fig:spect_pressure_JWST}
\end{figure}
\end{center}

\begin{center}
\begin{figure*}
\centering
\includegraphics[scale=0.35]{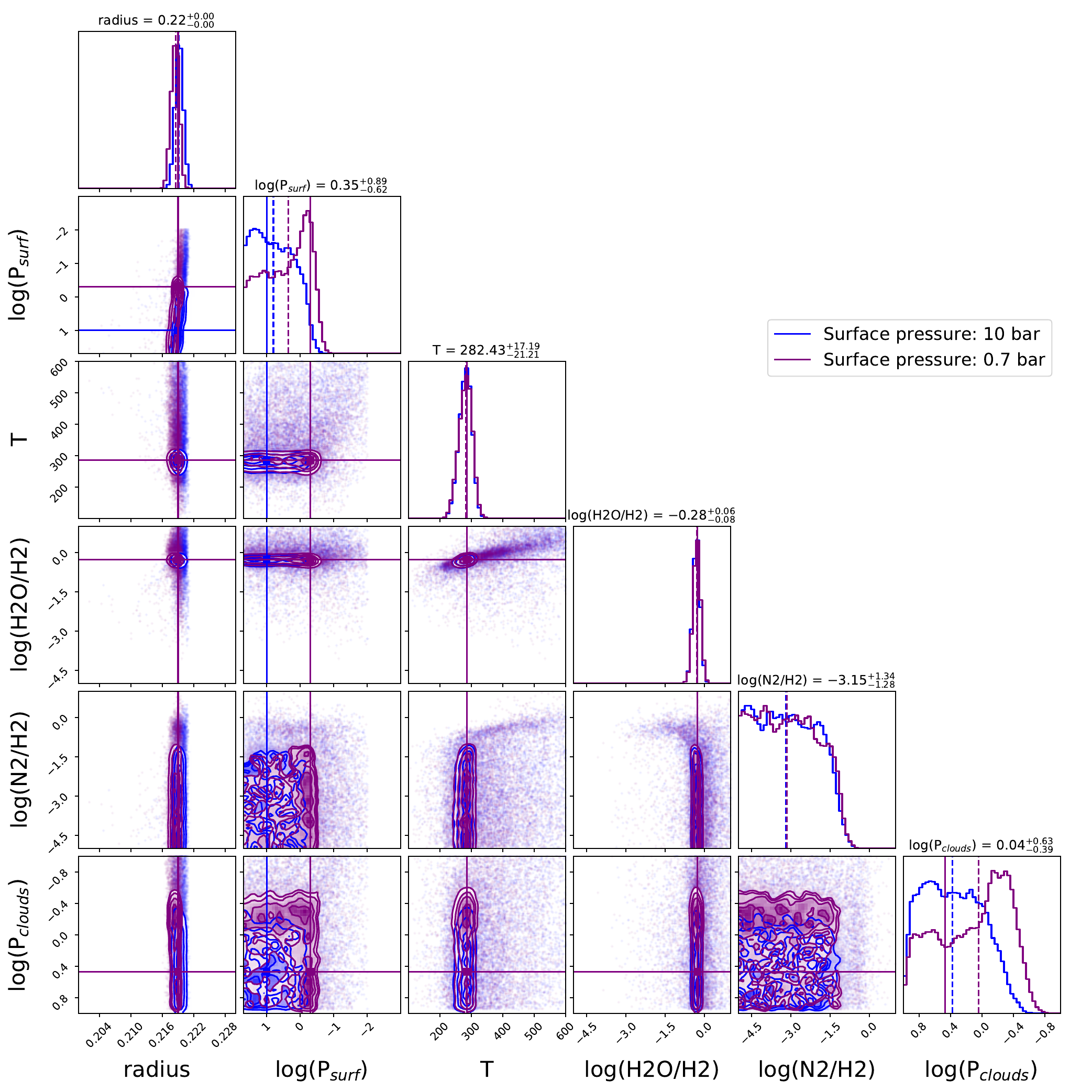}
\caption{
Posteriors for the retrievals where we attempt to recover the surface pressure. Blue: the forward model was using a surface pressure of 10 bar; Purple: the forward model was using a surface pressure of 0.7 bar. The forward models correspond to the ones in Figure \ref{fig:spect_pressure_JWST}.
}\label{fig:post_JWST_press}
\end{figure*}
\end{center}

Finally, we assess whether the surface pressure of K2-18\,b, if it exist, can be obtained from observations. From mass-radius considerations only, the presence and conditions at an hypothetical solid surface are unconstrained \cite{Tsiaras_k2-18,madhu_k218}. We simulated the cases of an atmosphere with a 10 bar, 1 bar and 0.7 bar surface pressures by varying the maximum pressure of our atmosphere grid. We plot the corresponding spectra in Figure \ref{fig:spect_pressure}. 
This test is important because determining the conditions at the surface is crucial to assess the potential existence of liquid or solid layers, and therefore to constrain the nature of the planet. We ran again the retrieval simulations from \cite{Tsiaras_k2-18} using different surface pressures. In the original study, the grid extended up to 10bar.  We find that all 3 surface pressures give the same results and would still be compatible with the observed WFC3 spectrum (see Figure \ref{fig:spect_pressure}) and confirm that current observations cannot determine conclusively the nature of the planet. We repeated the experiment for JWST simulated data and show three forward models, one for each surface pressure, for the heavy water solution 1 in  Figure \ref{fig:spect_pressure_JWST}.
Figure \ref{fig:spect_pressure_JWST} shows how for the same planet, the surface pressure influences the observed spectrum. While the observed spectra are different, the changes appear across the entire wavelength range, meaning that they should be very similar to changes in planet radius or cloud pressure. To investigate these degeneracies, we perform two retrievals for the cases with surface pressures of 10 bar and 0.7 bar and attempt to directly recover the pressure at the surface of our model (P$_{surf}$). We set the uniform priors for the retrieved surface pressure from 50 bar to 0.001 bar. The posterior distributions of these two retrievals is presented in Figure \ref{fig:post_JWST_press}.

From the posteriors in Figure \ref{fig:post_JWST_press}, one can see that the two retrievals provide very similar posterior distributions. The surface pressure seems to be difficult to constrain directly. The case with surface pressure of 10 bars is unambiguously converging towards a high surface pressure solution, characterised by a lower limit of around 1bar. The retrieval on the case with surface pressure of 0.7 bar does not provide a definitive answer as it presents two poorly separated modes (see posterior distribution). The first mode is a high pressure with clouds (pressure lower than 1 bar) solution, while the second solution is closer to the true forward model with low surface pressure (peaking at 0.7 bar) and no clouds. This suggests that there are some hints of the lower surface pressure in the simulated spectrum in the Rayleigh scattering part, the Collision Induced Absorption or in the cross sections pressure dependence. However, the characterisation of the surface pressure for cloudy super-Earth planets is likely to be difficult, even in the case of JWST and Ariel.

\section{DISCUSSION}

The results presented here suggest that observations with future space infrared observatories will allow to  characterise the nature of K2-18\,b. While current facilities may be limited to  disentangle between the three scenarios identified in \cite{Tsiaras_k2-18}, as well as to determine whether the planet has a liquid/solid surface or not, complementary observations may provide  additional constraints. 

\begin{itemize}
\item Constraining the Rayleigh/Mie slope: \\
Additional observations of K2-18b transits in the optical, i.e. 0.3-0.7 $\mu m$, could help to constrain the presence  of clouds or hazes in this atmosphere.  
 In this spectral region, the data may contain a lot of information concerning the planetary radius and the atmospheric scale height, which could be very informative. 
 \item Reflected light in eclipse observations or phase-curves: \\
 Eclipse observations in the optical could also help to identify and characterise clouds. 
 \end{itemize}

In general, clouds increase the planet albedo by reflecting visible light. \cite{Mansfield_albedo} has shown that cloud reflection could be distinguished from surface reflection by an increased albedo in the case of rocky planets. While this could be an interesting option, they only investigated surfaces for temperatures higher than 410K. They indeed highlighted the fact that water rich elements (formed at lower temperatures) could have a high albedo, which complicates the interpretation of the results. However, the reflected brightness of a primary atmosphere sub-Neptune or a planet with rocky surface should peak at phase 180$^{\circ}$ while an ocean world would have a peak brightness around phase 30$^{\circ}$ \cite{Zugger_ocean_scattering}, providing a direct method to separate these different scenarios. Using TauREx in forward mode, we investigated the thermal emission of K2-18\,b and found that the signal (flux ratio of the planet over the star: $F_p / F_s $) would be lower than 0.01 percent at 50 $\mu m$. This means that the emission spectrum of K2-18\,b is not observable with JWST. In the case of K2-18\,b, reflected light also presents huge challenges.  Indeed, the Signal strength for the reflected light case is $S_{ref} = \frac{A_g}{4} \left( \frac{R_p}{a}\right)^2$, where $A_g$ is the geometric Albedo and $a$ is the semi-major axis. Since K2-18\,b is orbiting far from its host star (a = 0.14 AU \cite{cloutier-k218_2017}), the reflected planet-to-star contrast remains too small ($S_{ref} \approx 4 \times 10^{-8}$ assuming an Albedo of 0.3) to be captured by current and next generation telescopes.  \\ \\
 

\section{CONCLUSIONS}

We simulated observations of the low gravity planet K2-18\,b as recorded with the next generation of space infrared observatories, i.e.  JWST and Ariel. 
K2-18\,b  is currently the only planet known in this regime with a confirmed water vapour detection.  The simulations were based on the 3 degenerate solutions identified in \cite{Tsiaras_k2-18} from the Hubble Space Telescope WFC3 observations: A Icy/Water world with significant water content in the atmosphere; A super-Earth with secondary atmosphere and trace water; A cloudy sub-Neptune with mainly primordial H$_2$/He. By performing a retrieval analysis of these scenarios, we show that the next generation of space telescopes will be able to distinguish among the 3 cases. While we choose the particular case of K2-18\,b for these simulations, our results demonstrate that next generation telescopes will significantly increase our understanding of the planets in the sub-Neptune desert. In the near future, observations of these worlds will allow to answer key open questions: What are their nature? Can super-Earth retain their primordial envelope? What are their formation and evolution history?  

In the case of K2-18\,b, the minimum required  observations vary from 2 combined transits with NIRISS and NIRSpec for JWST to 20 for Ariel. Increasing the number of observations inevitably leads to better constraint on the atmosphere of K2-18\,b. While the chemistry (both main gases and trace elements), temperature and clouds properties of K2-18\,b seems to be in reach of the future observatories, our retrieval simulations indicate that the surface pressure may be difficult to directly constrain. \\ \\

\vspace{5mm}
\section{Acknowledgements}

This project has received funding from the European Research Council (ERC) under the European Union's Horizon 2020 research and innovation programme (grant agreement No 758892, ExoAI) and under the European Union's Seventh Framework Programme (FP7/2007-2013)/ ERC grant agreement numbers 617119 (ExoLights). Furthermore, we acknowledge funding by the Science and Technology Funding Council (STFC) grants: ST/K502406/1, ST/P000282/1, ST/P002153/1, ST/S002634/1 and ST/T001836/1.



\clearpage
\newpage
\onecolumn

\bibliographystyle{spphys}
\bibliography{main}


\renewcommand{\floatpagefraction}{.9}%
\appendix 

\section*{Appendix: Additional figures}\label{Appendix_parameters}

\begin{center}
\begin{figure}[h]
\includegraphics[scale=0.5]{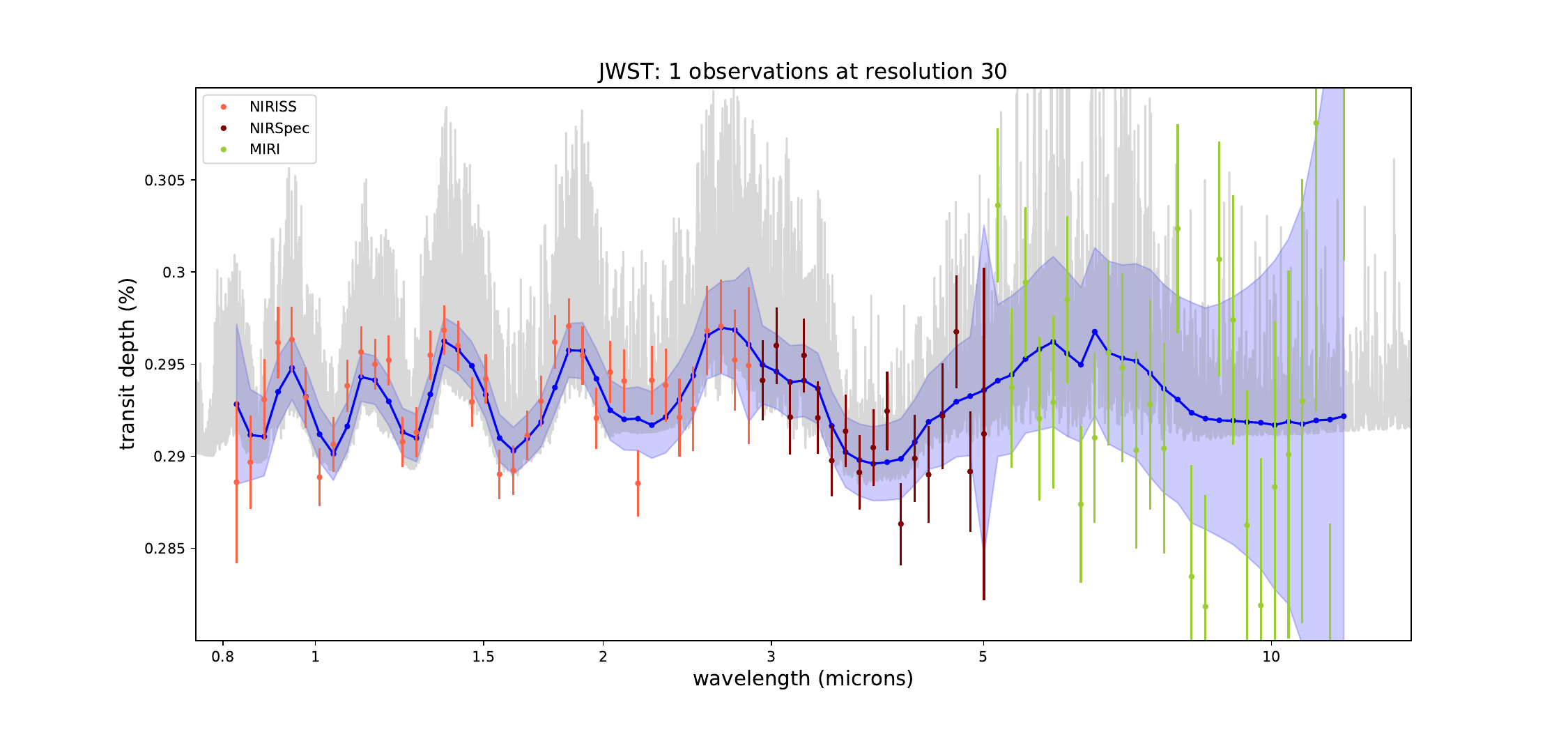}
\caption{
Observed spectra for the atmospheric scenario 1 with the different JWST instruments: NIRISS, NIRSpec and MIRI. The Error bars are displayed for a single transit with each instrument. The averaged model and 1$\sigma$ error are indicated by the shaded blue region.
}\label{fig:jwst_obs}
\end{figure}
\end{center}

\begin{center}
\begin{figure}[h]
\includegraphics[scale=0.6]{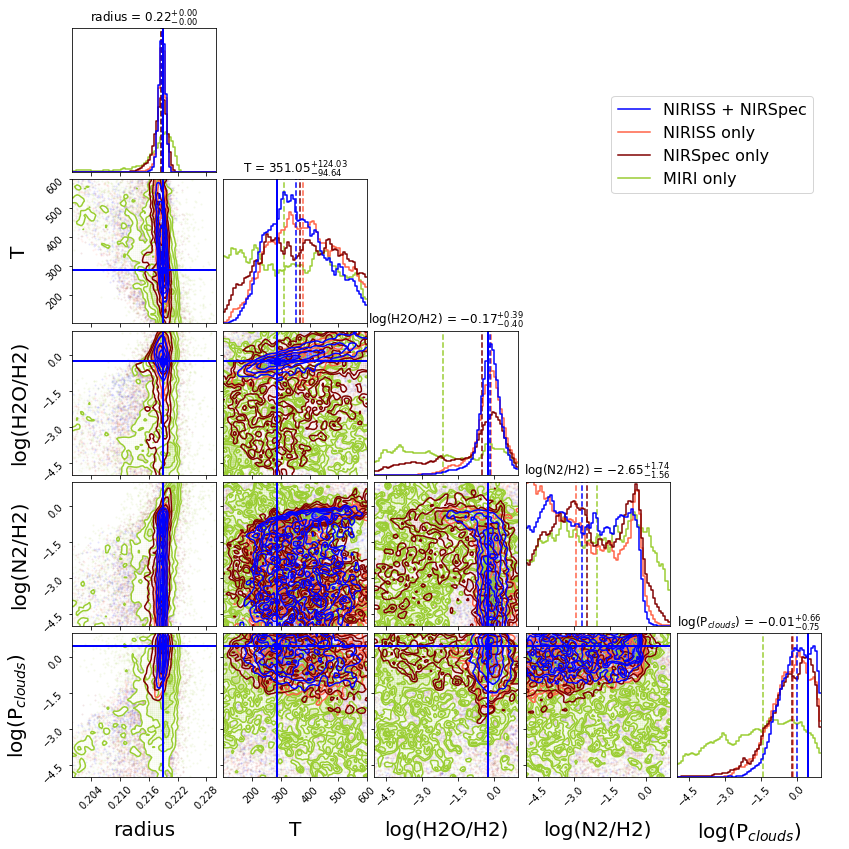}
\caption{
Retrieval posteriors for the atmospheric scenario 1 with different JWST setup. The NIRISS only scenario seems to provide similar performances than the NIRISS + NIRSpec case. If only NIRSpec is used, the water-to-hydrogen ratio is much more difficult to constrain, since only a single broad spectral modulation is present in NIRSpec wavelength coverage.
}\label{fig:jwst_inst}
\end{figure}
\end{center}

\begin{center}
\begin{figure}[h]
\includegraphics[scale=0.6]{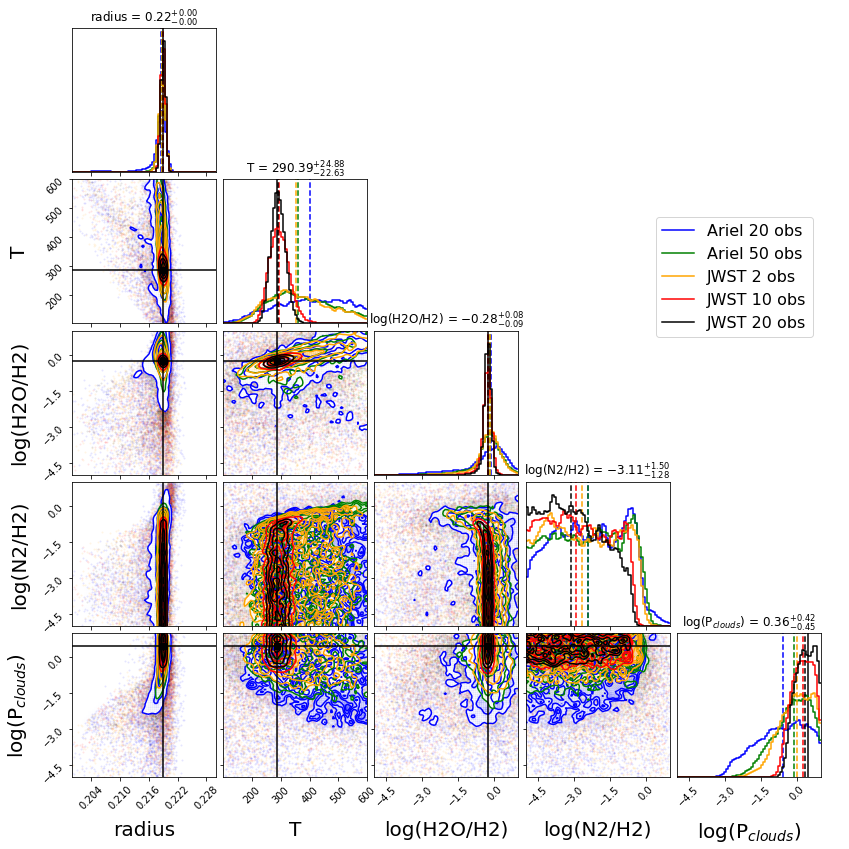}
\caption{
Retrieval posteriors for the atmospheric scenario 1: secondary atmosphere composed of H/He and H$_2$O.
}\label{fig:scenario1a}
\end{figure}
\end{center}

\begin{center}
\begin{figure}[h]
\includegraphics[scale=0.6]{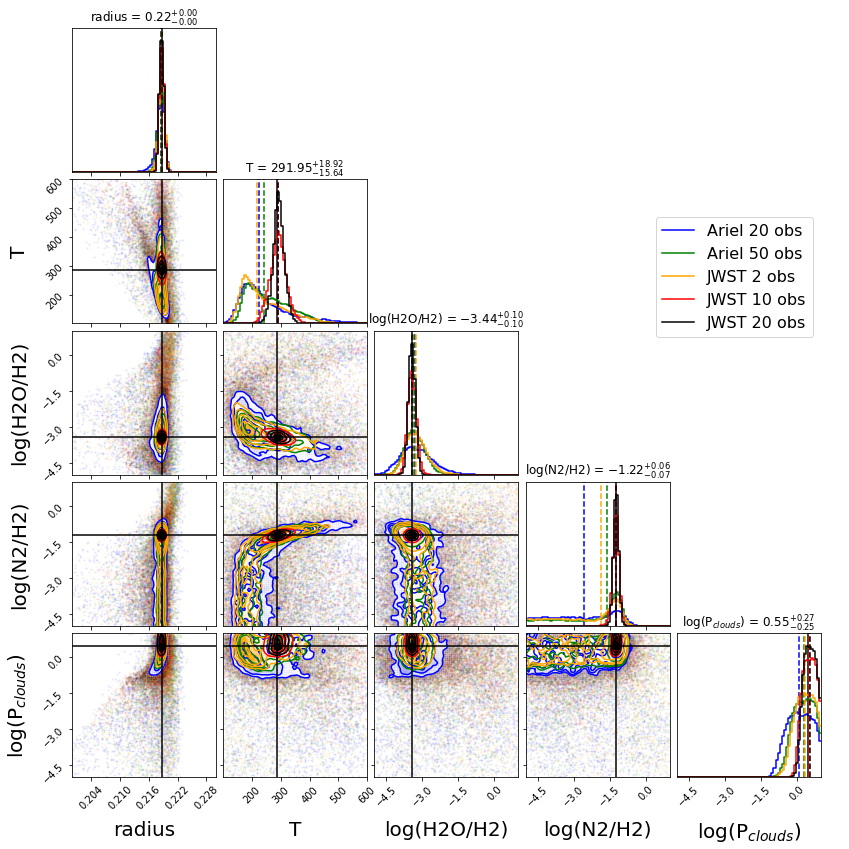}
\caption{
Retrieval posteriors for the atmospheric scenario 2: secondary atmosphere composed of H/He and another undetectable gas, i.e. N$_2$. Traces of H$_2$O are also present.
}\label{fig:scenario2}
\end{figure}
\end{center}

\begin{center}
\begin{figure}[h]
\includegraphics[scale=0.6]{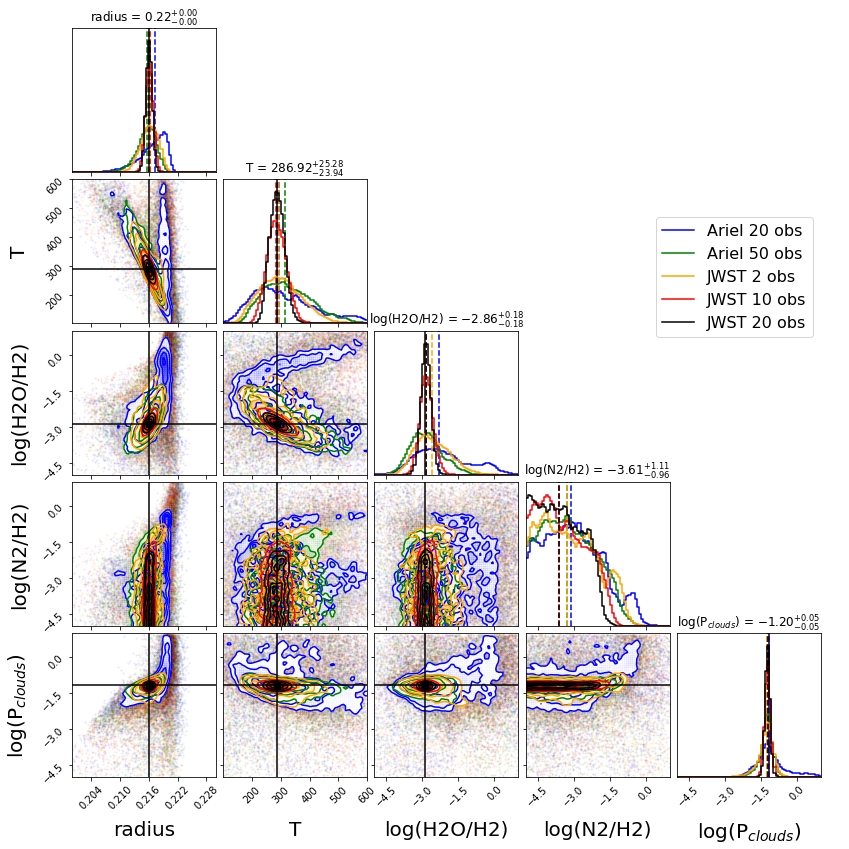}
\caption{
 Retrieval posteriors for the atmospheric scenario 3: primary atmosphere composed of H/He and clouds. Traces of H$_2$O are also present.
}\label{fig:scenario3}
\end{figure}
\end{center}

\begin{center}
\begin{figure}[h]
\includegraphics[scale=0.6]{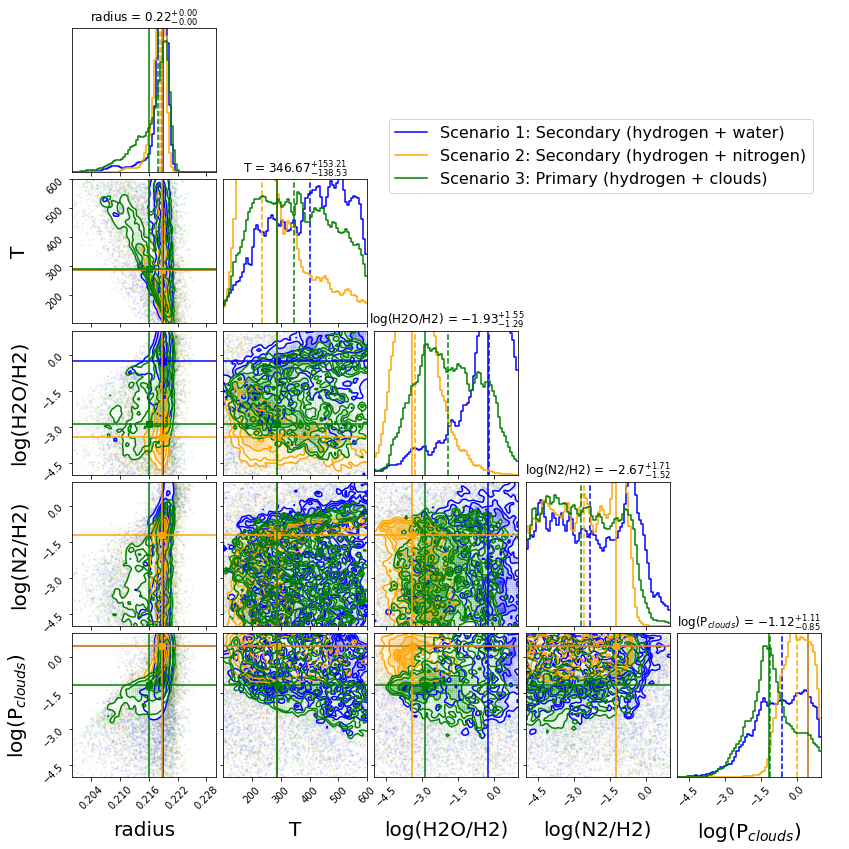}
\caption{
 Retrieval posteriors for the 3 scenarios in the case of 10 Ariel transits. The posteriors indicate a departure from a unique solution, however some of the atmospheric parameters are still correlated.
}\label{fig:post_ariel10}
\end{figure}
\end{center}



\end{document}